\documentclass{article}
\usepackage{arxiv}
\usepackage{moreverb}
\usepackage{graphicx}
\usepackage{textgreek}
\usepackage{booktabs}
\usepackage{amsmath,amssymb,amsfonts}
\usepackage[utf8]{inputenc}

\title{Using Near Infrared Spectroscopy and Machine Learning to diagnose Systemic Sclerosis}

\author{
  Joelle Feijó de França\\
  UNIBRA\\
  Recife\\
  Brazil\\
   \And
  Hugo Abreu Mendes \\
  POLI\\
  University of Pernambuco\\
  Recife \\
  Brazil \\
  \texttt{ham@poli.br}\\
  \AND
  Lucas Gallindo Costa \\
  Department of Electrical Engineering\\
  Federal University of Pernambuco\\
  Recife \\
  Brazil \\
    \AND
  Andrea Tavares Dantas \\
  Clinical Hospital\\
  Federal University of Pernambuco\\
  Recife \\
  Brazil \\
    \AND
  Angela Luzia Branco Pinto Duarte \\
  Clinical Hospital\\
  Federal University of Pernambuco\\
  Recife \\
  Brazil \\
    \AND
  Anderson Stevens Leônidas Gomes \\
  Physics Department\\
  Federal University of Pernambuco\\
  Recife \\
  Brazil \\
    \AND
  Emery Cleiton Cabral Correia Lins\\
  Department of Biomedical Engineering\\
  Federal University of Pernambuco\\
  Recife \\
  Brazil \\
}

\begin{document}
\maketitle
\newpage
\begin{abstract}
The motivation of this work is the use of non-invasive and low cost techniques to obtain a faster and more accurate diagnosis of systemic sclerosis (SSc), rheumatic, autoimmune, chronic and rare disease. The technique in question is Near Infrared Spectroscopy (NIRS). Spectra were acquired from three different regions of hand's volunteers. Machine learning algorithms are used to classify and search for the best optical wavelength. The results demonstrate that it is easy to obtain wavelength bands more important for the diagnosis. We use the algorithm RFECV and SVC. The results suggests that the most important wavelength band is at 1270 nm, referring to the luminescence of Singlet Oxygen. The results indicates that the Proximal Interphalangeal Joints region returns better accuracy's scores. Optical spectrometers can be found at low prices and can be easily used in clinical evaluations, while the algorithms used are completely diffused on open source platforms.
\end{abstract}


\footnotetext{\textbf{Abbreviations:} SSc, Systemic Sclerosis; NIR, Near Infrared; SVM, Support Vector Machine; RFECV, Recursive Feature Elimination with Cross-Validation; PDA, Photodetector Amplifier}

\section{Introduction}
Systemic sclerosis (SSc) is characterized as a chronic autoimmune disease consisting of a connective tissue disorder causing not only fibrosis of the skin, but also blood vessels and organs such as the lung, kidney, heart and gastrointestinal tract. This disease occurs more frequently in women, with a higher incidence in the age group between 30 and 50 years of age \cite{steen1990epidemiology, proudman2007pulmonary}. Several epidemiological studies indicate that african-descendant are more likely to have severe visceral manifestations than caucasians, as well as a worse prognosis and higher prevalence \cite{ho2003clinical, chifflot2008incidence}.

Collagen is an insoluble fibrous protein, the most abundant and that makes up a large part of the body's total protein mass. Its main function is to maintain the physical structure of tissues due to its mechanical resistance from its macro organization \cite{lee2001biomedical, shoulders2009collagen}. 
According to the manifestation of collagen dysfunction, systemic sclerosis can be classified as cutaneous diffuse and cutaneous limited. Diffuse SSc is identified by the rapid development of skin thickening symmetrically at the proximal and distal extremities, face and trunk, with a higher prevalence of visceral organ involvement. On the other hand, the limited cutaneous SSc is present only in distal extremity skin and on the face and, less frequently involves internal organs, but it can present CREST syndrome (calcinosis, Raynaud phenomenon, esophageal dysmotility, sclerodactyly, and telangiectasia).The diffuse deposition of this collagen causes dysfunction that results from the excessive stiffness of the tissue, thus compromising its functioning \cite{masi1980preliminary, avouac2011preliminary, van20132013}.
Progressive tissue fibrosis occurs by deposition of excessive amounts of collagen and other extracellular matrix proteins. This fact occurs due to persistent and uncontrolled fibroblast immune activation at various sites, increasing within the connective tissue, causing stiffness beyond expected for this tissue \cite{kuroda1997gene, neville2016patologia}.

The diagnosis of SSc at the present time is made through the observation of its clinical manifestations, such as Raynaud's phenomenon, in the laboratory targeting auto-antibodies, or even through tissue biopsy, however, because it is an invasive procedure, it is less used \cite{albilia2007small, allanore2015systemic, neville2016patologia}. This procedure currently makes the detection of the disease late, a fact that compromises the effective therapeutic intervention to the patient. Thus, research is being carried out to enable the early diagnosis of the disease with the objective of providing an effective treatment to the patient seeking improvement in their life quality.
\begin{figure}
\centering
\includegraphics[width=1.0\columnwidth]{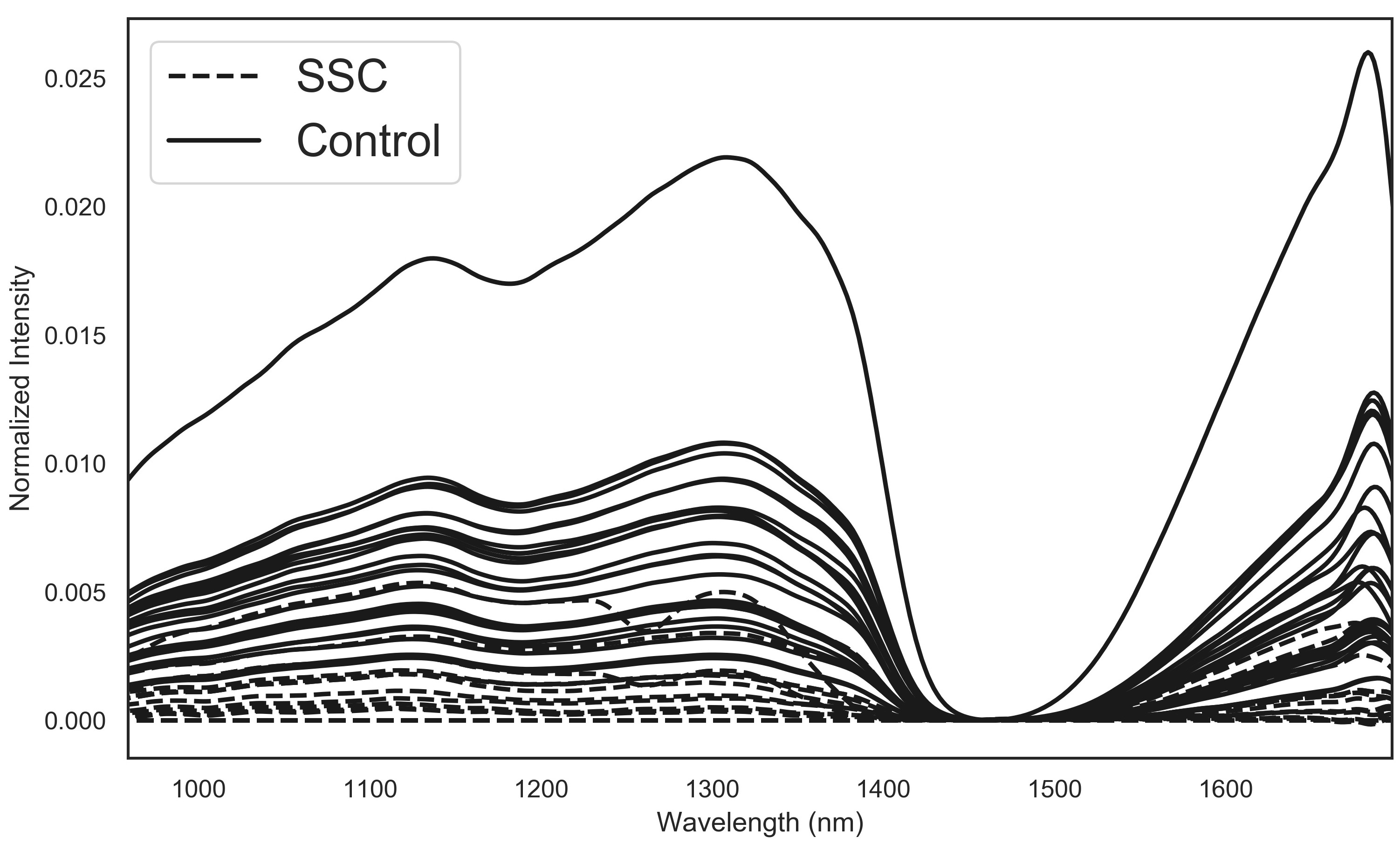}
\caption{SSc and Control Spectra, obtained on the Proximal Interphalangeal Joints region. Spectra already preprocessed. Wavelength axis in nanometers.}
\label{fig_1}
\end{figure}
In addition to the search for early SSc diagnosis, non-invasive methods are searched for in order to provide a better quality of life for the patient, due to this fact, optical techniques are promising because they do not use ionizing radiation and are capable of capturing information from the biological tissue. This uptake occurs according to the optical properties that the tissue presents, varying according to the distribution of its fundamental components \cite{proudman2007pulmonary}.

At present, researches that use optical properties are promising due to the behavior of the interaction of light in the biological tissue. In general, when an electromagnetic wave strikes a material medium, in addition to being able to penetrate the medium, this radiation can be reflected, absorbed and refracted. The probability of occurrence of each of these phenomena depends on the optical properties of the medium, which in turn depend on the wavelength of the radiation, thus the optical properties of biological tissue can be described through the coefficients of absorption, refraction, reflection, transmission, and scattering when it interacts with non-ionizing radiation \cite{tsai2013study, vo2014biomedical}. In this work, the methodology involves the study of absorbance spectroscopy using Near Infrared (NIR) radiation. This technology is based on the emission of low energy radiation, typically through optical fibers, over one or more locations on the surface of the tissues under investigation and measurement of back-scattered intensities \cite{hielscher2002near, scholkmann2014review}.

\begin{figure}
\centering
\includegraphics[width=1.0\columnwidth]{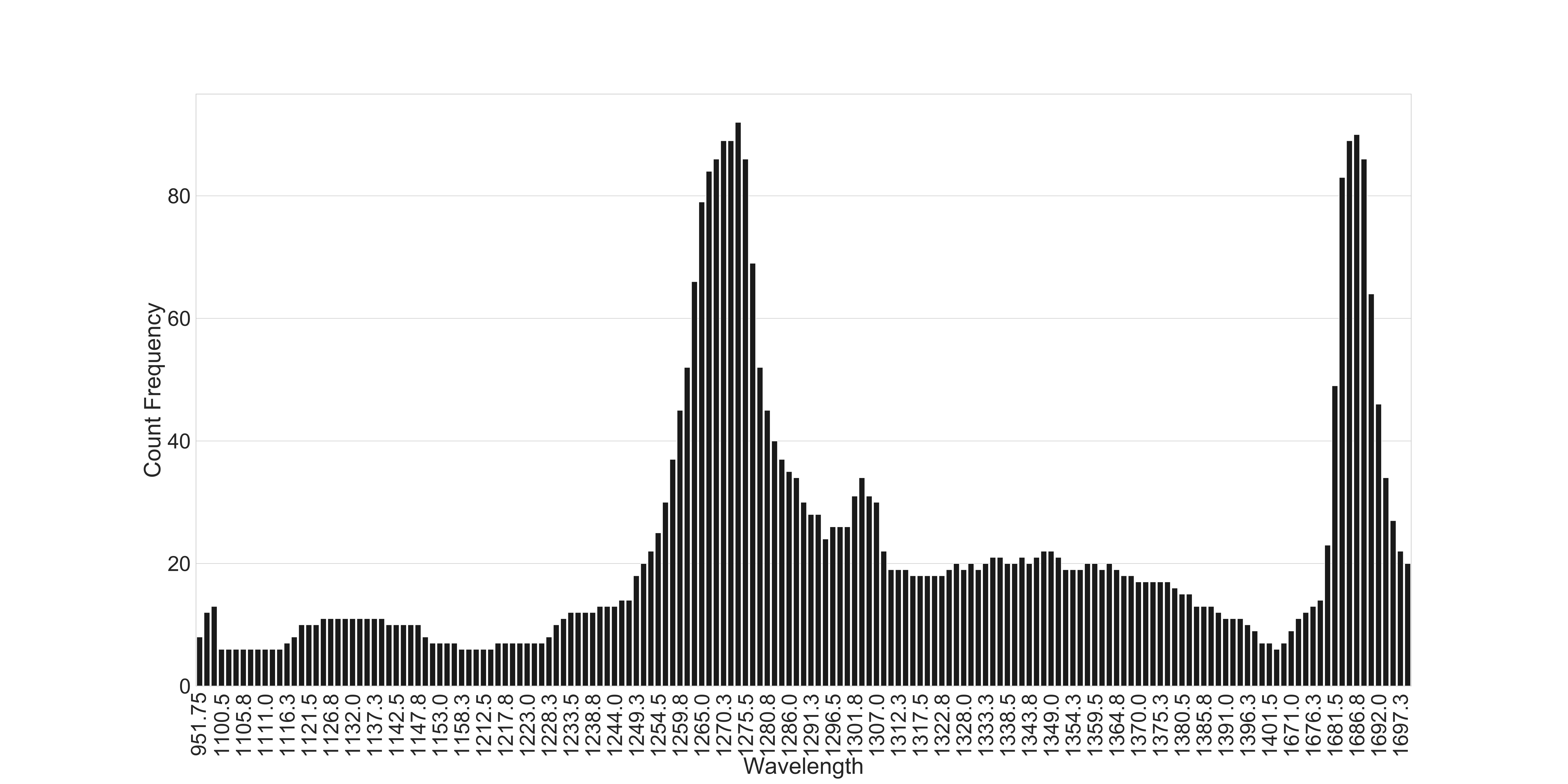}
\caption{histogram of wavelengths chosen for classification in the Proximal Interphalangeal Joints collection region. For a hundred SVC training and test executions. Wavelength in nanometers.}
\label{fig_2}
\end{figure}
\section{Methods}
Electromagnetic waves with wavelengths in the infrared range are easily absorbed by human tissue due to the large amount of water present in their composition, whose spectrum favors the absorption of light in that region. On the other hand, visible light, which occupies a range of wavelengths between approximately 400 nm and 700nm, is strongly scattered throughout human tissue. Between the visible spectrum and the infrared region we have the near infrared, located between approximately 900nm to 1700nm. In this region, the NIR radiation is poorly absorbed by human tissue. Its scattering coefficient is such that it causes the incident light to spread in all directions, making it possible to detect this kind of radiation even after penetrating a few centimeters into the tissue. It is this optical window of human tissue that is exploited through a form of spectroscopy, called NIRS \cite{vo2014biomedical}.

The near infrared absorbance spectroscopy study for characterization of groups as being healthy or diseased is discussed in follow of this section. Some techniques of digital signal processing, statistics and machine learning will be discussed, which, in all, serve as an argument for the diagnosis and characterization of the main markers of the disease, from an optical spectroscopy point of view, using near infrared. The instrumentation for obtaining the spectra will also be discussed in this section.

The experiment was characterized by a near-infrared spectroscopy study with diffuse infrared radiation in the human skin: a group with 25 patients clinically diagnosed as absent from the manifestations of Scleroderma (control group) and another group with 17 patients clinically diagnosed as symptomatic of Scleroderma. The spectra were obtained at three different anatomical regions of each participant's hand: Proximal Phalanges, Metacarpal, and Proximal Interphalangeal Joints.

The demonstration of the experiment conducted in people, exemplified in this paper, has the approval of the Ethics and Research Committee Involving Humans, Federal University of Pernambuco (CEP UFPE). According to the Code of Ethics established by resolution number 466/12 of the National Health Council of the Ministry of Health, for experiments in humans. CAAE: 66030917.2.0000.5208, protocol number: 2,185,420. Date of approval: 07/2017.
\subsection{Spectra acquiring and preprocessing}
\begin{figure}
\centering
\includegraphics[width=1.0\columnwidth]{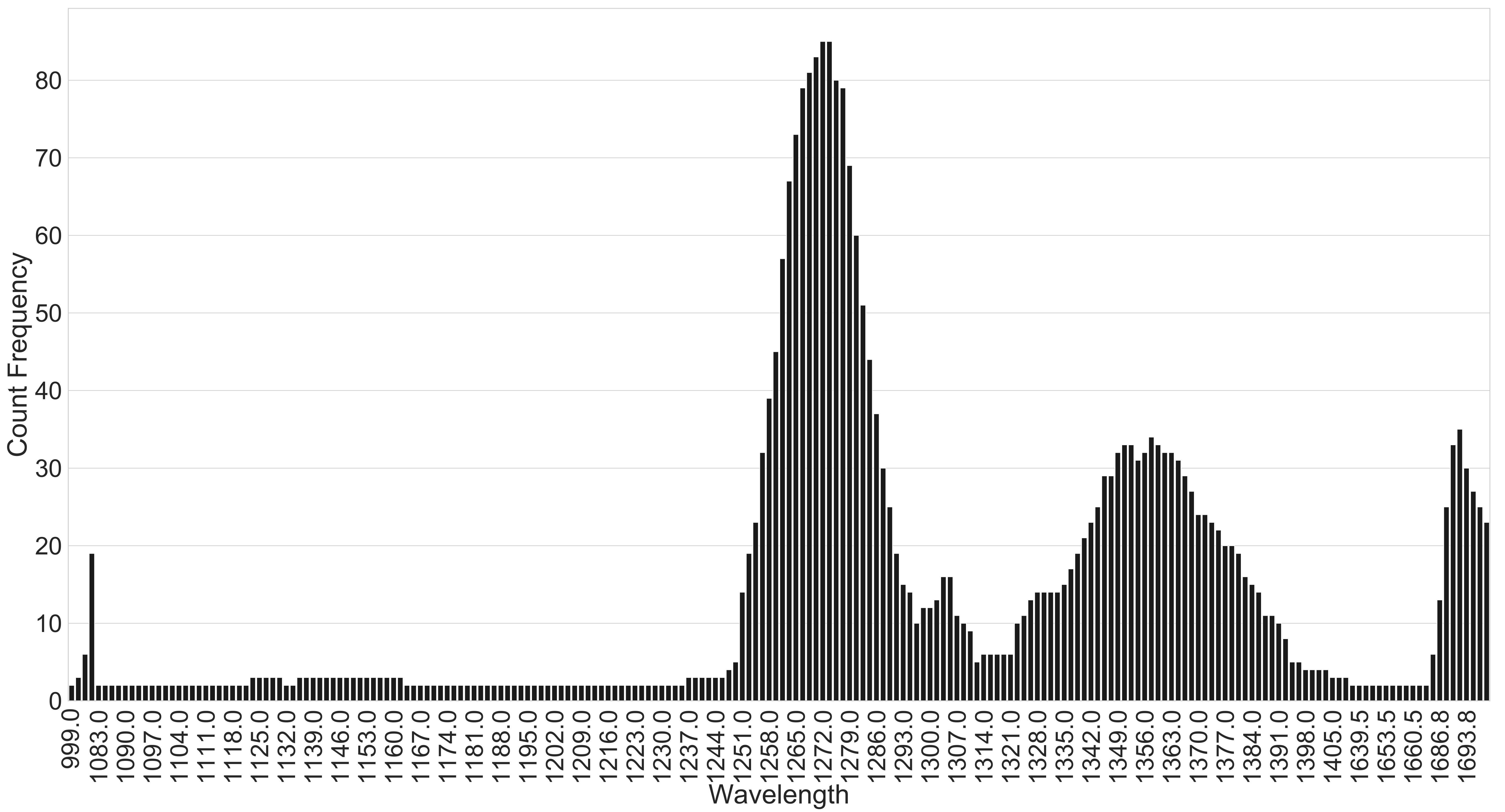}
\caption{histogram of wavelengths chosen for classification in the Metacarpal collection region. For a hundred trainings and tests, of the classification. Wavelength in nanometers.}
\label{fig_3}
\end{figure}
A DSR-CXR-512 digital spectrometer (StellarNet, Inc., USA) was used to collect the data. This equipment consists of two modules that operate in different spectral regions. It has a concave grid sensor to compose UV, VIS, NIR spectra (200nm - 1000nm), while the other sensor has an InGaAs PDA (Photodetector Amplifier) in order to compose NIR spectra (900nm - 1700nm). Despite the composition of the spectrometer, data were collected only in the near infrared range. The radiation source is an SL1 Tungsten Halogen, from the same manufacturer, with 200 W/m\textsuperscript{2} and 2800K color temperature and a wide spectral Range of 350-2500nm.

The spectroscopy data obtained by the measurement in a clinical setting were processed using Python3 programming language and its popular packages NumPy and Pandas. Overall, as an open source programming language Python is constantly increasing its popularity among data scientists, engineers and researchers in several areas. Besides, software written in Python is well compatible with both Linux and Windows OS (operational systems) while other available packages can easily bind to better scalable systems, stored in the cloud. All those features justify its use and perfect applicability for this work.

A model is then needed to explain the relationship between baseline, noise and observed signal. In \cite{giguere2015optimization} there is a simple method of baseline and noise correction presumes that the intensity values observed are the sum of the true signal, a reference signal, and noise, as shown by 
Equation \ref{eq_1}.
\begin{equation}
    y = y_{r}+b+\epsilon
\label{eq_1}
\end{equation}

Where y is the signal obtained, y\textsubscript{r} is the real signal, \textit{b} is the baseline and \textepsilon{} is white noise. A rectangular window is used, which functions as a moving average filter, aiming to filter the \textepsilon{} component of the signal. To set the spectral data in a unity scale and in order to recreate a linear system, the baseline of the raw data had to be removed. To remove the baseline, the spectra were subtracted from the minimum value, so that each spectrum had at least one of its components at zero intensity. With this accomplished, then normalization by the radiation source was done, in order to obtain the absorbance spectra, by dividing each intensity value of collecting spectra by the correspondent in wavelength from the radiation source intensity specter.

Figure ~\ref{fig_1}shows definitively how the spectra of the two groups to be studied remained. It is visible, that the two groups possess distinctive mean observance levels. However, the two groups are not visible completely separable. The separability will be formally proven using the machine learning techniques proposed within the scope of this paper, and better described in the next subsection.

For the purpose of balancing quantities of samples between the control group and the SSc group, false samples were created, so the number of 25 samples was established for each of the two groups.
\begin{figure}
\centering
\includegraphics[width=1.0\columnwidth]{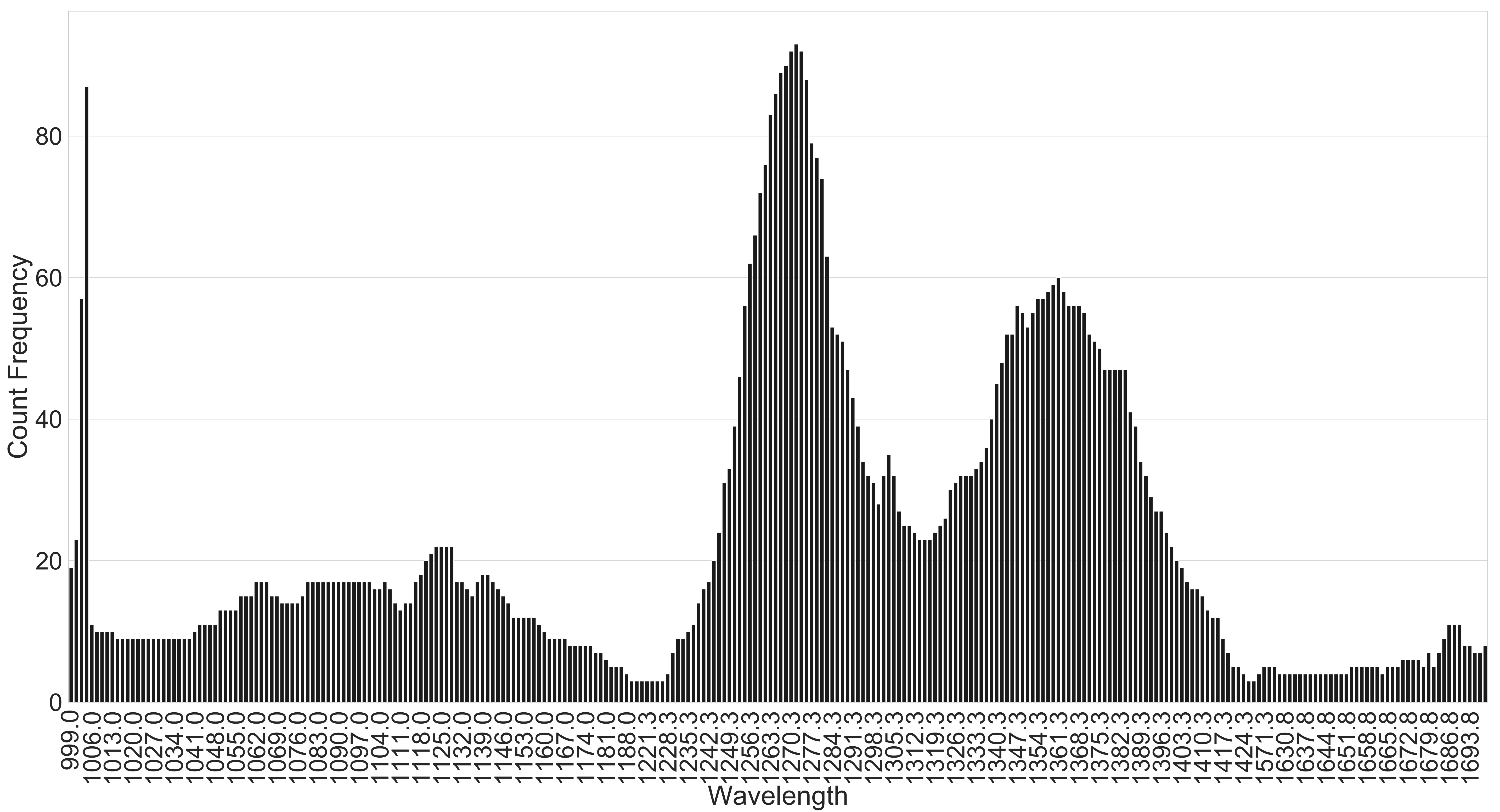}
\caption{histogram of wavelengths chosen for classification in the Proximal Phalanges region. For a hundred trainings and tests, of the classification. Wavelength in nanometers.}
\label{fig_4}
\end{figure}
\subsection{Machine Learning}

With the spectra ready for analysis, an approach was taken to determine how good a binary classification would be between the two groups. More specifically, an SVC (Support Vector Classification). In \cite{burges1998tutorial} there is a vast explanation on the theory of the SVM (Support Vector Machine) algorithm. For the implementation of this algorithm, we used the Sklearn package available for Python3 \cite{scikit-learn}. The Sklearn package implements the algorithm based in another library, called LIBSVM \cite{chang2011libsvm}, implementing as well the use of probabilistic outputs for support kernels \cite{platt1999probabilistic}.

A grid search was performed, being an exhaustive search method for specific parameters for the classifier algorithm. From the exhaustive search, it was obtained that the algorithm should have a linear kernel and have a penalty parameter of 10\textsuperscript{3} order of magnitude.

In addition to the hyper-parameterization of the classifier, a feature selection, with recursive elimination and cross-validation, also available in Sklearn was made, in order to find the best number of attributes and which are these. Besides the value of the information alone, this methodology avoids overfitting \cite{guyon2002gene}. The algorithm learns to select only the most important wavelengths to obtain greater accuracy in the classification task between the SSc and control groups.

For each voluntary collection region, the SVC algorithm with RFECV (Recursive Feature Elimination with Cross-Validation) was run 100 times, which is a good number of executions when dealing with heuristics solutions \cite{carling2016statistical}. The input information is the spectra of a random selection of individuals from the two groups, as training data, with the percentage of test data of 20\%. The algorithm adapts to each training sample, returning which wavelengths were optimally chosen to better classification. With the complementary percentage of individuals, the accuracy is obtained on the test data. The next section will present the results and discussion.
\section{Results}
\begin{figure}
\centering
\includegraphics[width=0.75\columnwidth]{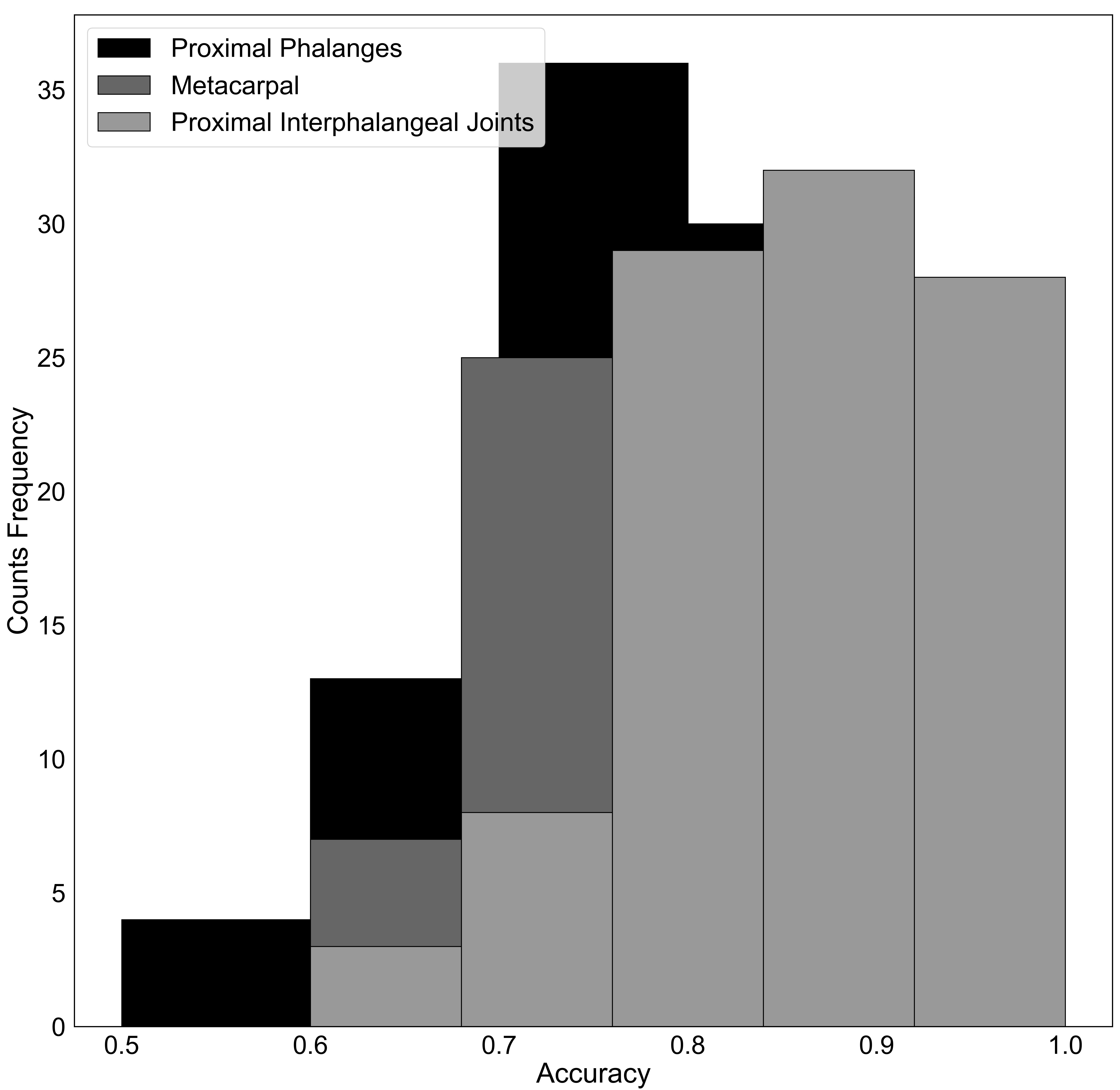}
\caption{Accuracy histogram for all spectra collection region and run times.}
\label{fig_5}
\end{figure}

\begin{table}
    \caption{Rounded most important bandwidth means for classification per region in nanometers} 
    \centering
    \begin{tabular}{llll}
        \toprule 
        Hand Skin Region & First band & Second Band & Third Band \\
        \midrule
        Proximal Interphalangeal Joints & 1270 & 1350 & 1685\\
        Metacarpal & 1270 & 1360 & 1690\\
        Proximal Phalanges & 1270 & 1360 & -\\
    \end{tabular}
\label{Tab_01}
\end{table}
Figures ~\ref{fig_2}, ~\ref{fig_3} and ~\ref{fig_4}show in sequence the results for each spectra collection region. In the figures wavelength axis, some values are omitted, for better visualization. Table ~\ref{Tab_01}presents the wavelength bands found to be the most important for the classification task. It is worth noting that the importance of the band is directly linked to the number of times it has been counted among the 100 executions of the algorithm. Finally, Figure ~\ref{fig_5}, shows the accuracy distribution result for each region. U-tests were performed to better establish which regions was most statistically accurate. For all tests the hypothesis of one distribution being statically greater than other was analysed, all \textrho{} values resulted were greater than 0.99. The Proximal Interphalangeal Joints region is the one with the best representation of accuracy, while the Proximal Phalanges is the worst. 

\section{Discussion}
The most frequent wavelength band is 1270 nm, for this case, in \cite{baier2007direct} there is an explanation about the detection by the luminescence of Singlet Oxygen (\textsuperscript{1}O\textsubscript{2}) in human skin and cells. Those reactive oxygen species are generated by the absorption of UVA radiation in endogenous molecules present in cells, damaging its proteins \cite{vile1995uva, davies2003singlet, glaeser2011singlet}.

Singlet Oxygen has been shown to elicit an increase in collagenase in human fibroblasts \cite{scharffetter1993singlet}. It is also demonstrated that its behavior within these cells is highly localized, but its high life span may serve as the basis for its diffusion in larger scale \cite{jimenez2008kinetics}. On the introduction of \cite{herrick2000double}, the pathogenesis of SSc is discussed, taking into account the importance of oxygen-free radicals as contributors to tissue damage. In \cite{ahsan2003oxygen} work, the relation of all reactive oxygen species and autoimmune systemic pathologies are well debated. As the purpose of the work is to argument on the use of machine learning algorithms and NIR spectroscopy over the SSc diagnoses issue, it is left for the reader to search in the cited references more about the SSc pathogenesis.

From 1000 nm water is now one of the most absorbed components in the skin, along with fat \cite{scholkmann2014review, wilson2015review}. No relevant references were found that singularly distinguish specific components at 1360 or 1690 nm. 
\section{Acknowledgements}
Thanks to the members of the Laboratory of Multidisciplinary Engineering of the Federal University of Pernambuco and to the professors and professionals of the Clinic Hospital of the same university.

\bibliographystyle{unsrt}
\bibliography{refs}
\end{document}